\documentclass [11pt]{article}
\usepackage{amsmath,amssymb,epsfig,cite,color,verbatim}
\usepackage{graphics,float}
\usepackage{overpic} 
\usepackage{mathrsfs}
\usepackage{slashed,bbold}
\usepackage{soul}

\numberwithin{equation}{section}

\title{The fate of chiral symmetry in two-flavor matrix adjoint QC$_2$D}
\date{\today}
\author{Nirmalendu Acharyya$^1$\footnote{nirmalendu@iitbbs.ac.in}, Prasanjit Aich$^{2}$\footnote{prasanjita@iisc.ac.in}, Sayan Bhakta$^{2}$\footnote{sayanbhakta@iisc.ac.in}, \\
Ranita Mudi$^1$\footnote{a22ph09014@iitbbs.ac.in} \, and Sachindeo~Vaidya$^2$\footnote{vaidya@iisc.ac.in} \\
${}^1${\small School of Basic Sciences, Indian Institute of Technology Bhubaneswar, Jatni, Khurda, Odisha 752050, India}\\
${}^2${\small Centre for High Energy Physics,  Indian Institute of Science, Bengaluru, 560012, India}\\
}
\begin{document}

\maketitle
\vspace*{-1cm} 
\begin{abstract}
In the matrix model of two-flavor adjoint QC$_2$D, we study the low-lying states and their properties in the intermediate-to-strong (Yang-Mills) coupling regime. The model has a classical $SU(2)_R$ chiral symmetry and the eigenstates of the Hamiltonian can be organized in its irreps. We construct the energy eigenstates in presence  of a chiral chemical potential $c$ using the variational techniques.  We find that when $c=0$, the ground state is always a $SU(2)_R$ singlet, irrespective of the coupling strength $g$. However, as $g$ is tuned from intermediate to strong coupling, the system under goes a crossover from a unique to a doubly degenerate ground state. The degeneracy in the strong coupling regime spontaneously breaks the axial $\mathbb{Z}_8$, while preserving the chiral symmetry.  

When $c \neq 0$, we find that there can be level crossings which correspond to quantum phase transitions (QPTs). Depending on the ground state, there are three possible phases. The $SU(2)_R$ symmetry is spontaneously broken in only one of these phases, and this phase can only emerge for intermediate $g$ with moderate values of $c$. In the $g-c$ plane this phase corresponds to a narrow window, outside which $SU(2)_R$ is always preserved. 

\end{abstract}

\section{Introduction}

Four-dimensional gauge theories with adjoint matter provide a fertile testing ground for non-perturbative dynamics, in the context of both particle physics and condensed matter systems \cite{Hietanen:2009az,Bursa:2009we, DelDebbio:2010hu,  DeGrand:2011qd, Catterall:2007yx, Hietanen:2008mr, Catterall:2008qk, Bergner:2016hip, Bergner:2017gzw, Poppitz:2019fnp, Catumba:2024jau,Sarkar:2022rti}. Among these, the two-flavor adjoint QC$_2$D (i.e. the $SU(2)$ gauge theory with two adjoint Weyl quark flavors) is of particular interest \cite{Hands:2000ei,DelDebbio:2009fd,Athenodorou:2014eua,Athenodorou:2021wom, Bennett:2022bhc, Das:2025utp, Bi:2018xvr, Bi:2019gle, Cordova:2018acb,Anber:2018iof,Wan:2018djl}. In the chiral (massless quarks) limit, the system has a continuous global $SU(2)_R$ chiral symmetry and an axial $U(1)_A$ (classical) symmetry. In the quantum theory, the $U(1)_A$  is broken by the axial anomaly, leaving behind a $\mathbb{Z}_8$. The residual symmetry structure provides powerful constraints on the possible IR phases of the theory \cite{Cordova:2018acb}.  Hence this theory is an important testing ground for exploring fundamental questions about confinement, chiral symmetry breaking, and conformal fixed points.

Despite its apparent simplicity, the long-standing debate regarding the nature of the infrared dynamics of the two-flavor adjoint QC$_2$D remains unresolved. There are two different theoretical predictions: one suggests that the theory is confining and breaks its chiral symmetry spontaneously, while the other says that the theory flows to an interacting conformal fixed point in the IR, i.e., it lies within the conformal window. This debate has been the central subject of several investigations, with arguments rooted in effective field theory analyses \cite{Bi:2018xvr, Anber:2018iof, Wan:2018djl}, or deformations of $\mathcal{N}=2$ supersymmetric Yang-Mills theory \cite{Cordova:2018acb}, or first-principles lattice simulations  \cite{Athenodorou:2014eua, Bi:2019gle, Athenodorou:2021wom,Bennett:2022bhc}. However, the lattice simulations are  performed at finite quark mass and require  delicate chiral extrapolations, which are notoriously difficult to control near a putative conformal fixed point. The effective field theory descriptions, in turn, depend on specific assumptions about the relevant IR degrees of freedom and are not uniquely dictated by the symmetries alone. These  limitations seem to preclude a definitive consensus, leaving open the fate of the $SU(2)_R$ symmetry, and the exact phase structure.

Here, we consider a matrix model of two-flavor adjoint QC$_2$D and study its low-lying energy eigenstates with the aim to understand the $SU(2)_R$ chiral symmetry.  The matrix model (proposed in \cite{Balachandran:2014iya} for pure Yang-Mills theory and \cite{Pandey:2016hat} coupled to quarks)  is a (0+1)-dimensional (quantum mechanical) approximation of the non-Abelian gauge theory.  Despite this dramatic simplification, it succeeds in capturing the intriguing features such as the twisted nature of the gauge bundle \cite{Singer:1978dk, Narasimhan:1979kf,Balachandran:2014iya} and axial anomaly \cite{Acharyya:2021egi}. As computations here are relatively straightforward, it can be an effective probe for the low energy dynamics.  The matrix model estimates of the glueball and hadron spectra in  $SU(3)$ gauge theory  \cite{Acharyya:2016fcn, Pandey:2019dbp} demonstrate remarkable accuracy and act as the benchmarks to extend it to uncharted territories.  The subsequent investigations have confirmed certain features of QC$_2$D and  revealed  several novel ones \cite{Pandey:2016hat, Acharyya:2017uhl,  Acharyya:2024pqj, Acharyya:2026uhx, Acharyya:2026rvx}. All these results encourage us to  explore  the fate of chiral symmetry in  two-flavor adjoint QC$_2$D.  We find that in absence of the chiral chemical potential, the chiral symmetry remains intact in the intermediate-to-strong coupling $g$ regime. A non-zero chiral chemical potential $c$ can induce a $SU(2)_R$-broken phase, but for a  very narrow $g-c$ window.

The Hamiltonian of this model has $SU(2)_R$ chiral symmetry, and  the states can be organized in its irreps. Using  variational methods, we construct the color-singlet energy eigenstates transforming in different representations of $SU(2)_R$. We restrict our attention to the intermediate-to-strong coupling regime ($0.4 \lesssim g \lesssim 5$), where the convergence of the numerical data is excellent. Too far outside this regime, the convergence is poor with standard variational methods and quantitative results become unreliable.

Of course, the weak and the ultra-strong coupling regimes are very interesting in their own right. But preliminary investigations indicate that there is a phase transition at weak coupling. Similarly as $g\to \infty$, the potential develops flat directions, again requiring its own careful investigation.  These lead to loss of convergence of the numerical data at both limits.  To regain the conceptual and analytic control, we need to use bigger trial sets and perform careful finite size scaling.  In any case,  the results at these two ends (whatever they may be)  do not overturn the findings in our regime of interest $0.4 \lesssim g \lesssim 5$.

When $c=0$, we find that the ground state  is a spin-0 $SU(2)_R$ singlet in both the intermediate and strong coupling regimes, implying that the $SU(2)_R$ chiral symmetry remains intact. In the intermediate coupling regime, the ground state is unique and hence the system has unbroken axial $\mathbb{Z}_8$ symmetry. However, as the coupling strength increases, there is a crossover to a doubly degenerate (both spin-0 $SU(2)_R$ singlets) ground state in the strong coupling regime.  The degenerate states have different quark content and spontaneously  break the  axial $\mathbb{Z}_8$  to $\mathbb{Z}_4$. Remarkably, these findings  are in agreement with the results from the full field theory \cite{Anber:2018iof}.

In contrast, when  $c\neq 0$, the degeneracy of the ground state in the strong coupling regime is lifted. Now, there are level-crossings in the ground state as we tune $g$.   At any such crossing, the ground state changes abruptly, leading to a quantum phase transition (QPT). Different phases have different ground states and several observables like the quark number, $SU(2)_R$ Casimir etc.  jump discontinuously at the transition points.     Among these phases, there is one with a $SU(2)_R$ triplet ground state:  in this phase,  the chiral symmetry is spontaneously broken. This phase emerges  only when $c$ and $g$ lie in a particular window, whose boundaries we establish numerically.    
Further, we show that at the phase boundaries (level crossings),  the axial-$\mathbb{Z}_8$ is spontaneously broken to either a $\mathbb{Z}_2$ or a $\mathbb{Z}_4$.

This article is organized as follows. In section \ref{sec_2}, we present the Hamiltonian of the model.  We consider the intermediate-to-strong coupling regime and construct the low-lying energy eigenstates transforming in the different irreps of $SU(2)_R$ and compare their energies. The results are presented in section  \ref{sec_3}. We discuss the properties of the ground state  in absence of  chiral chemical potential in section \ref{sec_3_1}. Subsequently  in section \ref{sec_3_2}, we demonstrate the level crossings in presence of the chiral chemical potential and QPTs.  We study the properties of the phases and the phase boundaries and show that for suitable choice of $g$ and $c$, the chiral symmetry is spontaneously broken. We end with a discussion in section \ref{sec_4}.

\section{Hamiltonian and Symmetries} \label{sec_2}
 The construction of the matrix model involves in a fundamental way the pullback of the Maurer-Cartan form of $SU(2)$ (or $SU(N)$) to the spatial three-sphere $S^3$ of radius $R$ \cite{Narasimhan:1979kf, Balachandran:2014iya}. The phase space for the glue dynamics is generated by the matrix $M_{ia}$ ($i,a=1,\ldots,3$) and their conjugate momenta $P_{ia}=-i \frac{\partial}{\partial M_{ia}}$ . We introduce two-flavor adjoint quarks $b_{\alpha a f}$ ($\alpha=1,2$ and  $f=1,2$), which are  time-dependent Grassmann-valued matrices \cite{Pandey:2016hat}. The Hamiltonian for this model (minimal coupling) is
\begin{eqnarray}
H&=&\frac{1}{R} \Big( \frac{1}{2} P_{ia} P_{ia} + \frac{1}{2} M_{ia} M_{ia} - \frac{g}{2} \epsilon_{ijk}\epsilon_{abc} M_{ia} M_{jb} M_{kc} + \frac{g^2}{4} \epsilon_{abc}\epsilon_{ade} M_{ib} M_{jc} M_{id} M_{je} + \nonumber \\
&& \quad i g \epsilon_{abc} b_{\alpha b f}^\dagger  \sigma^{i}_{\alpha\beta} b_{\beta c f} M_{ia} +c Q_0\Big)\label{Ham_1}
\end{eqnarray}
where $Q_0\equiv b_{\alpha a f}^\dagger b_{\alpha a f}$ is the total quark number and $R^{-1}$ determines the energy scale.

 The physical Hilbert space $\mathcal{H}_{phys} $ is the set of colorless states (which are annihilated by the Gauss law): 
\begin{eqnarray}
G_a | \phi \rangle =0, \quad  | \phi \rangle \in \mathcal{H}_{phys}. 
\end{eqnarray}
where $G_a$'s are the generators of the Gauss law: 
\begin{eqnarray}
 G_a\equiv - \epsilon_{abc} M_{ib} P_{jc} - i \epsilon_{abc} b^\dagger_{\alpha bf} b_{\alpha c f},  \quad \quad [G_a, G_b]=i \epsilon_{abc} G_c. 
\end{eqnarray}

\textit{Chiral symmetry:}  For two flavors, we can define 
 \begin{eqnarray}
Q_p \equiv \frac{1}{2} b^\dagger_{\alpha a f}\tau^p_{ff'} b_{\alpha af},  \quad\quad [Q_p, Q_q]=i \epsilon_{pqr} Q_r \quad\quad p,q,r=1,2,3, \label{SU_2B_charges}
 \end{eqnarray}
 where $\tau^p$'s are the Pauli matrices. The operators  $\{Q_p: \, p=1,2,3\}$ generates a $SU(2)_R$.  It is easy to see that $[H, Q_p]=0$  for $p=1,2,3$ and hence, the Hamiltonian has a $SU(2)_R$ chiral symmetry. Here, $Q_3$ is related to difference of the number of quarks of two flavors  and may be identified as the ``baryon number".

 The charge $Q_0$ generates axial rotations $U(1)_A$ and is a classical symmetry. However, this symmetry is broken quantum mechanically  to a $\mathbb{Z}_8$  \cite{Acharyya:2021egi}. Although $Q_0$ does not generate a symmetry in the quantum theory, it remains a respectable observable.

The $SU(2)_R$ symmetry on the other hand survives in the quantum mechanical version.  However, we must entertain the possibility of its spontaneous breaking for some values of $g$ and $c$. In particular, if the ground state is a $SU(2)_R$ triplet, then this symmetry is spontaneously broken.

\section{Results}\label{sec_3}

\begin{figure}
\begin{center}
\includegraphics[width=16cm]{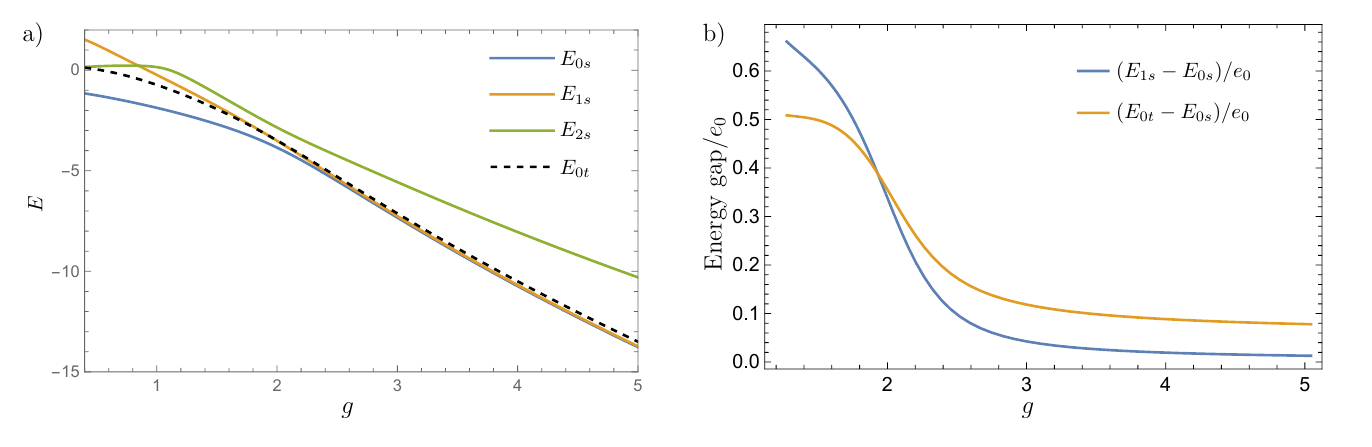}
\caption{a) Energies (in units of $R^{-1}$) of the lightest three $SU(2)_R$ singlets ($n=0,1,2$), and the lightest $SU(2)_R$ triplet  ($n=0$). b) Energy gaps $(E_{1s}-E_{0s})/e_0$ and $(E_{0t}-E_{0s})/e_0$ vs $g$, where $e_0 \equiv E_{2s}-E_{0s}$. }\label{Fig_2}
\end{center}
\end{figure}

For the numerical simulations, we expand the bosonic wavefunctions in the basis of harmonic oscillator energy eigenstates and truncate the expansions at a maximum number of oscillator quanta $N_b$ \cite{Acharyya:2024pqj, Acharyya:2026uhx}. We progressively increase $N_b$ until convergence is achieved in the energy eigenvalues. We find that the numerical estimates of the energies converge well with $N_b \sim 22$: the results for  $N_b=22$ and $N_b=25$ are practically indistinguishable.
 
 To characterize the ground state, it suffices to study the low-lying spin-0 states transforming as  $SU(2)_R$ singlets and  triplets. Eigenstates of higher representations and non-zero spin have much higher energies for any $g$ and $c$.

\subsection{$c=0$}\label{sec_3_1}  
The energies of the low-lying spin-0 $SU(2)_R$ singlets and triplets, denoted by $ E_{ns} $ and $E_{nt}$ respectively ($n=0,1,2 \ldots$) are shown in Fig.~\ref{Fig_2}a. As evident from the figure, the ground state across our regime of interest, with energy $E_{0s}$, is a $SU(2)_R$ singlet. Interestingly, in the regime $g \gtrsim 3.5$, the energy levels $E_{0s}$, $E_{0t}$, and $E_{1s}$ come very close to each other. Naturally, one has to ask whether these states become degenerate. These energies (in units of $R^{-1}$) are visibly very large compared to the energy gaps between them. To investigate the question of degeneracy, it is more natural and appropriate to compare these energy gaps to $e_0 \equiv E_{2s}-E_{0s}$, which is always non-zero.

The energy gaps $E_{0t}- E_{0s}$  and  $E_{1s}- E_{0s}$  in units of $e_0$ are shown in Fig.~\ref{Fig_2}b.   Both gaps decrease with increasing $g$.  Interesting, in the intermediate coupling regime, we observe a level  crossings between $E_{0t}$ and $E_{1s}$ at $g \simeq 2 $:   the first excited $SU(2)_R$ singlet is lighter than the lightest triplet for $g\gtrsim 2$. In the strong coupling regime ($g \gtrsim 3.5$),  we find that  $(E_{1s}- E_{0s})/e_0$ becomes approximately zero, while $(E_{0t}- E_{0s})/e_0$ has a small but non-zero value.

Based on these observations, we can draw the two following inferences for the $c=0$ case:
\begin{enumerate}
\item[i)]  the lightest $SU(2)_R$ triplet is heavier than the ground state for all $g$, and the chiral symmetry is unbroken.
\item[ii)] in the intermediate coupling regime, the ground state is a unique $SU(2)_R $ singlet, while as the coupling strength increases,  there is crossover to a two-fold degenerate ground state (both $SU(2)_R$ singlets). 
\end{enumerate}
We will provide further evidence for these two observations below.

\begin{figure}
\begin{center}
\includegraphics[width=16cm]{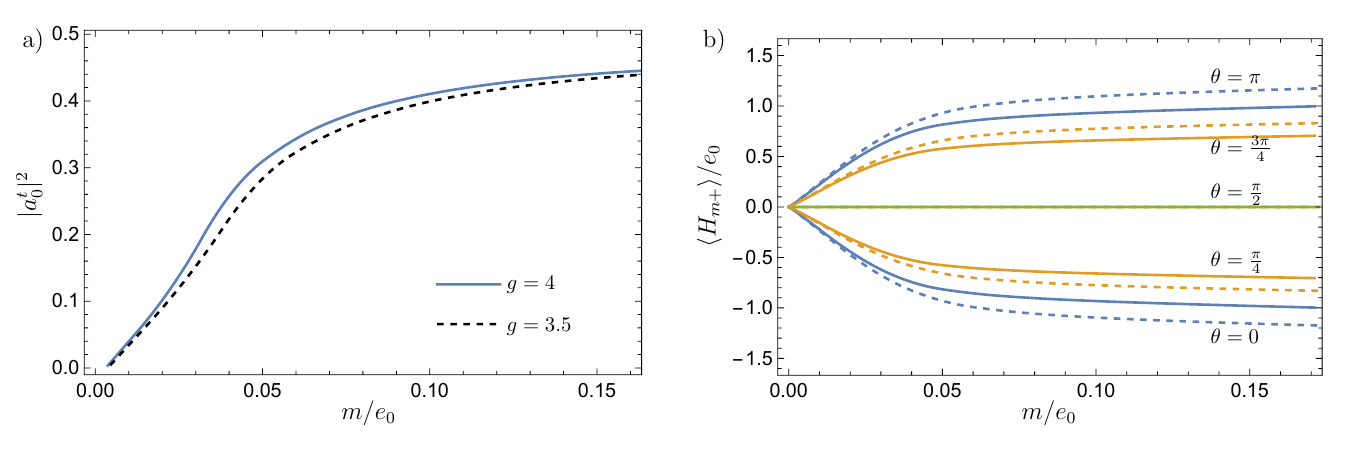}
\caption{ a) $|a_0^{t}|^2$ vs $m/e_0$ in the strong coupling regime for any $\theta$.  b) $\langle H_{m+}\rangle/e_0$ vs $m/e_0$ in the strong coupling regime for various $\theta$. The solid lines correspond to $g=4.0$ and the dashed lines correspond to $g=3.5$. }\label{Fig_3}
\end{center}
\end{figure}

More information about the nature of the ground state can be obtained by adding the mass term to the Hamiltonian. As we will show below, this will provide further evidence that the ground state in the massless  limit is indeed a $SU(2)_R$ singlet. 

The mass term is $m (\cos \theta H_{m +} + i \sin \theta H_{m -})$: 
\begin{eqnarray}
H_{m\pm} \equiv b_{\alpha a 1}^\dagger \sigma^2_{\alpha \beta} b_{\beta a 2}^\dagger \pm b_{\alpha a 2}\sigma^2_{\alpha \beta}  b_{\beta a 1},
\end{eqnarray}
with $m \in \mathbb{R}_0^+$ and $\theta \in[0,2\pi)$. It is straightforward to see that $[Q_3, H_{m\pm}]=0$ but $ [Q_{1,2}, H_{m\pm}] \neq 0$. Hence $H_{m\pm}$ breaks the $SU(2)_R$ symmetry explicitly to $U(1)_B$.

For any non-zero $m$ and arbitrary $\theta$, the ground state $|\Psi_{gs} \rangle$ has $Q_3=0$ (i.e is a ``meson'') and can be expanded as
\begin{eqnarray}
|\Psi_{gs} (m, \theta) \rangle = \sum_{n=0}^\infty \Big(a_{n}^{s} |\Psi_{ns}\rangle 
+ a_{n}^{t} |\Psi_{nt}\rangle + \ldots\Big),  \quad\quad a_{n}^{s}, a_{n}^{t}\ldots \in \mathbb{C}, 
\end{eqnarray}
where $\{|\Psi_{ns}\rangle, |\Psi_{nt}\rangle\ldots\}$ are eigenstates (with $Q_3=0$) of the Hamiltonian with $m=0$, each of which transforms in a fixed irrep of $SU(2)_R$. The coefficients $\{ a_{n}^{s}, a_{n}^{t}\ldots\}$ depend on $m$ and $\theta$.  As the matrix elements of $H_{m, \pm}$ are non-zero between states in different  irreps of $SU(2)_R$, the ground state degeneracy, if any, in the massless case is lifted by $H_{m \pm}$.

For any $g$, $m$ and $\theta$,  the contribution of  $|\Psi_{0t}\rangle$ in the ground state  can be quantified via 
 
\begin{eqnarray} 
|a_{0}^{t} |^2= |\langle\Psi_{0t}|\Psi_{gs}(m, \theta)\rangle|^2
\end{eqnarray}
which is, in general, non-zero. In the  strong coupling regime ($g \gtrsim 3.5$), the dependence of  $|a_{0}^{t} |^2$  on $m$ for different $\theta$ is shown in Fig.~\ref{Fig_3}a. For fixed $g$, we find that $|a_{0}^{t} |^2$ steadily decreases with decreasing $m$ and vanishes in the $m \to 0$ limit for any $\theta$. Further, the ground state expectation value $\langle H_{m+}\rangle_{gs}\equiv \langle \Psi_{gs}|H_{m+}| \Psi_{gs} \rangle$ (Fig.~\ref{Fig_3}b) remains non-zero for  $m>0$ but 
 
\begin{eqnarray}
\lim_{m \to 0 }  \langle H_{m+} \rangle_{gs}  \simeq 0. 
\end{eqnarray}
The vanishing  $|a_{0}^{t} |^2$ and $ \langle H_{m+} \rangle_{gs}$ both reaffirm that  the ground state in the $m\to 0$ limit is a $SU(2)_R$ singlet. 
 \begin{figure}
\begin{center}
\includegraphics[width=18cm]{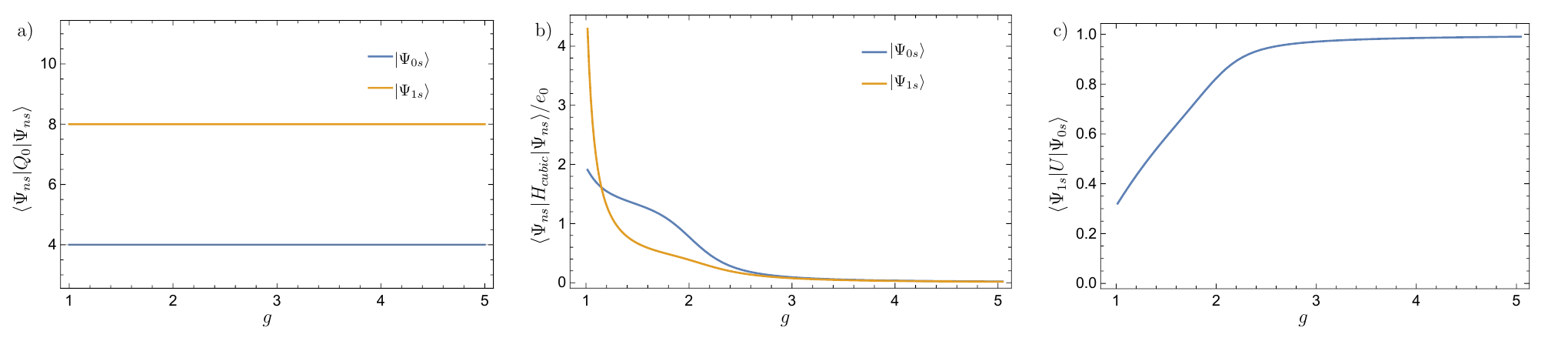}
\caption{a) $\langle  \Psi_{ns}| Q_0 |\Psi_{ns}\rangle$ vs $g$.  b)The matrix element of the cubic term $H_{cubic}$ in the $|\Psi_{0s}\rangle$ and $|\Psi_{1s}\rangle$  states as a function of $g$. c) $\langle \Psi_{1s}| U |\Psi_{0s}\rangle$ vs $g$.  }\label{Fig_4}
\end{center}
\end{figure}

We now turn to the second observation: in the strong-coupling regime, the two lightest $SU(2)_R$ singlet states, $|\Psi_{0s}\rangle$ and $|\Psi_{1s}\rangle$, become degenerate and form a two-fold degenerate ground-state manifold. For a low-energy effective description, it is sufficient to truncate the full Hilbert space to this two-level subsystem, and we shall focus on the effective Hamiltonian acting within it. 

For all values of $g$, we find $\langle  \Psi_{0s}| Q_0 |\Psi_{0s}\rangle \simeq 4$ and $ \langle  \Psi_{1s}| Q_0 |\Psi_{1s}\rangle \simeq 8$ (Fig.~\ref{Fig_4}a). So $|\Psi_{0s}\rangle$ contains on average 4 quarks, while $|\Psi_{1s}\rangle$ contains on average 8 quarks. Since the maximum quark number in the system is 12, the 8-quark state can equally be viewed as a 4-hole state. Particle and hole states are related by a transformation $U$
 \begin{eqnarray}
 U: b_{\alpha a f} \to \sigma^2_{\alpha \beta} b_{\beta a f}, \quad M_{ia} \to - M_{ia}.  \label{U_trans}
 \end{eqnarray}
If the energies of the two states are identical,  i.e. $\langle \Psi_{0s}| H |\Psi_{0s}\rangle = \langle \Psi_{1s}| H |\Psi_{1s}\rangle$,  then $U$ is a symmetry of the effective Hamiltonian for this two-level subsystem.

If $U$ is a symmetry, then the Hamiltonian of the two-level subsystem must commute with it.  However, it is straightforward to see that $U$ does not commute with the cubic interaction $H_{cubic} \equiv  \frac{1}{2} \epsilon_{ijk}\epsilon_{abc} M_{ia} M_{jb} M_{kc}$.  Therefore, for the transformation $U$ to be a symmetry, the cubic term must vanish in this doublet subspace (off-diagonal entries vanish because of different quark content):
\begin{eqnarray}
\langle \Psi_{0s}| H_{cubic} |\Psi_{0s}\rangle =\langle \Psi_{1s}| H_{cubic} |\Psi_{1s}\rangle = 0. 
\end{eqnarray}
 Furthermore, these degenerate wavefunctions are related by $|\Psi_{1s}\rangle =U |\Psi_{0s}\rangle$ and thus  must satisfy 
\begin{eqnarray}
\langle \Psi_{1s}| U |\Psi_{0s}\rangle = 1.   
\end{eqnarray}

In Fig.~\ref{Fig_4}b, we show the numerically obtained values of $\langle \Psi_{0s}| H_{cubic} |\Psi_{0s}\rangle$ and $\langle \Psi_{1s}| H_{cubic} |\Psi_{1s}\rangle$ as function of $g$. In the intermediate coupling regime, both  are non-zero and their values differ from each other. However, as $g$ increases, they decrease and in the strong coupling regime, they  become (practically) zero. 

On the other hand, $\langle \Psi_{1s}| U |\Psi_{0s}\rangle$ is plotted as a function of $g$ in Fig.~\ref{Fig_4}c. In the intermediate coupling regime, its value remains small and increases steadily with $g$.  As $g$ approaches the strong coupling regime,  $\langle \Psi_{0s}| U |\Psi_{1s}\rangle \to 1$.

These strongly suggests that the ground state in the strong coupling regime is doubly degenerate, consisting of two spin-0 $SU(2)_R$ singlets. There are further implications of these findings. The axial anomaly breaks the classical $U(1)_A$ symmetry to a residual $\mathbb{Z}_8$. If $|\Psi_{0s}\rangle $ (with $Q_0=4$) and $|\Psi_{1s}\rangle $ (with $Q_0=8$) have the same energy, then so does any linear combination of them. However, such a linear combination is invariant only under an axial $\mathbb{Z}_4$ transformation. Consequently, the degenerate ground state in the strong coupling regime spontaneously breaks the residual $\mathbb{Z}_8$ axial symmetry to a $\mathbb{Z}_4 \subset \mathbb{Z}_8$. Remarkably, this exact nature of the strong coupling ground state and symmetry structure  has been predicted in the full field theory framework by Anber and Poppitz \cite{Anber:2018iof}. 

In summary, we find that in the intermediate coupling regime $(g\lesssim 3.5)$, the ground state is a unique $SU(2)_R$ singlet and the residual axial symmetry is $\mathbb{Z}_8$. In the strong coupling regime, there is a crossover to a doubly degenerate ground state.  Both of  these states are  $SU(2)_R$ singlet, but they break the axial $\mathbb{Z}_8$ to $\mathbb{Z}_4$.

 \begin{figure}\begin{center}
\includegraphics[width=18cm]{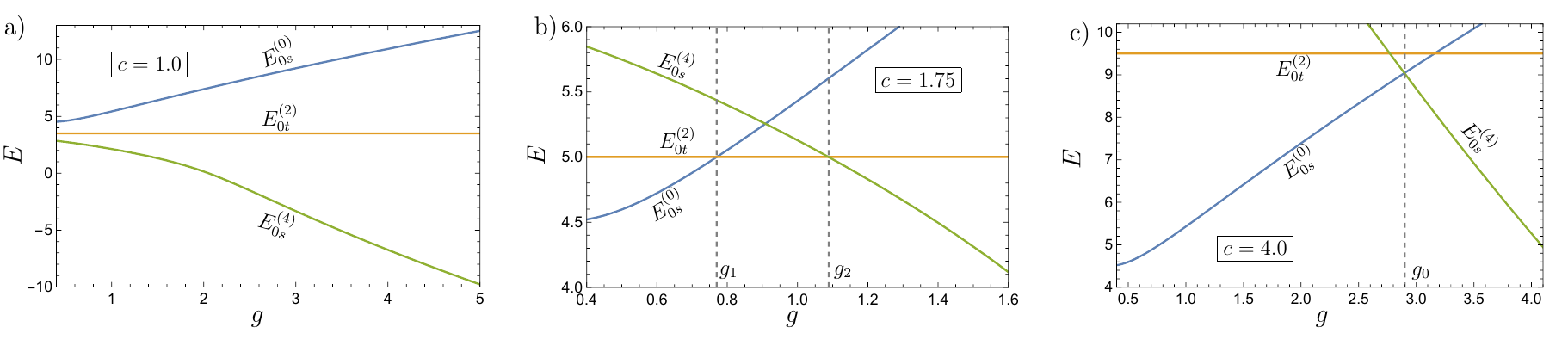}
\caption{ $E_{0s}^{(4)}$, $E_{0s}^{(0)}$ and $E_{0t}^{(2)}$ (in units of $R^{-1}$) vs $g$ for different $c>0$. a) When $c=1.0 <\Delta_{24,min}$, there is no level crossing.  b) When $\Delta_{24,min}<c= 1.75<\Delta_{24}(g_\ast)$, there are two level crossings at $g_1$ and $g_2$.  c) When $c=4.0>\Delta_{24}(g_\ast)$, there is a single level crossing at $g_0$.   } \label{Fig_sing_trip_crossing}
 \end{center}
\end{figure}

\subsection{Non-zero chiral chemical potential}\label{sec_3_2}
When $c>0$,  the states get an additional contribution $c \langle Q_0\rangle$ and we will denote their energies as $E_{n\ldots}^{(\langle Q_0\rangle)}(g,c)$. This dependence on $c$ opens the interesting possibility that for a fixed $g$, varying $c$ can change the ground state because of level crossings (i.e. QPTs). Each phase is characterized by a distinct ground state $|\Psi_{gs}(g,c)\rangle$ which has energy $E_{gs} (g,c)$ and its quantum numbers.

 For $c=0$,  we saw in the last section that the ground state is $|\Psi_{0s}\rangle$  for $g\lesssim 3.5$, while it is the degenerate pair  ($|\Psi_{0s}\rangle, |\Psi_{1s}\rangle$) for $g\gtrsim 3.5$.  As $\langle Q_0\rangle_{0s}\simeq 4$ and $\langle Q_0\rangle_{1s}\simeq 8$, it is not difficult to see that  a non-zero $c$ lifts this degeneracy. Thus if $c$ is sufficiently small and positive, then $|\Psi_{0s}^{(4)}\rangle$ is the unique ground state for all $g$ with energy $E_{0s}^{(4)}$. Further, for positive $c$, no state with $\langle Q_0 \rangle >4$ can ever be the ground state. On the other hand if $c$ is sufficiently large, states with $\langle Q_0 \rangle <4$ may have energy less than $E_{0s}^{(4)}$. In particular, we find that  on fixing $c$ to some positive value, the lightest singlet with $\langle Q_0\rangle=0$, or the lightest triplet with $\langle Q_0\rangle=2$ may be the ground state for some range of $g$. 
 
Indeed we do find such level crossings (see Fig.~\ref{Fig_sing_trip_crossing}).  
  \begin{figure}\begin{center}
\includegraphics[width=16cm]{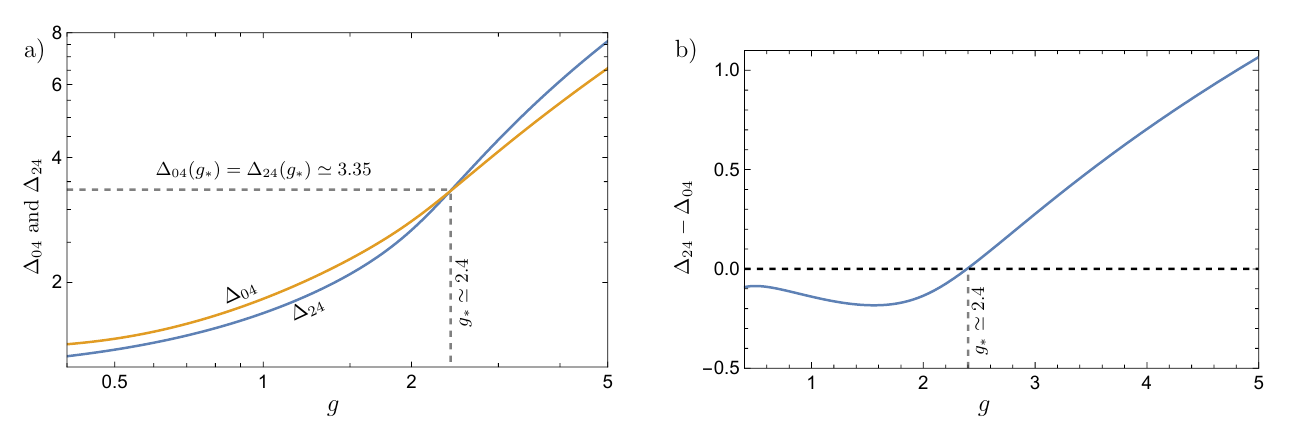}
\caption{  a) $\Delta_{24}$ and $\Delta_{04}$ as functions of $g$. b)  The difference $\Delta_{24}-\Delta_{04}$ as functions of $g$.  $\Delta_{24}-\Delta_{04}$ changes sign at $g_{\ast} \simeq 2.4$, denoted by dashed vertical lines. At $g_\ast$, $\Delta_{24}=\Delta_{04} \simeq 3.35$, denoted by the dashed horizontal line in panel a).  } \label{Fig_delta_comb}
 \end{center}
\end{figure}
To determine the conditions for such level crossings, we introduce the quantities:
\begin{eqnarray}
\Delta_{24} \equiv \frac{1}{2}\Big[E_{0t}^{(2)}(g,0)-E_{0s}^{(4)}(g,0)\Big], \quad\quad \Delta_{04} \equiv \frac{1}{4}\Big[E_{0s}^{(0)}(g,0)-E_{0s}^{(4)}(g,0)\Big]. 
\end{eqnarray}

Irrespective of the value of $g$, both $\Delta_{24}$ and $\Delta_{04}$ are positive (see Fig.~\ref{Fig_delta_comb}a), but their difference $(\Delta_{24}-\Delta_{04})$ does not have a definite sign (see Fig.~\ref{Fig_delta_comb}b). 
As evident from the figure, $\Delta_{24}$ and $\Delta_{04}$ are increasing functions with their minima at $g \simeq 0.4$, which we denote as $\Delta_{24, min} \simeq 1.3$ and $\Delta_{04, min} \simeq 1.5$.  
Further, the difference $(\Delta_{24}-\Delta_{04})$ vanishes at $g_\ast \simeq 2.4$, with $\Delta_{24}(g_\ast) \simeq \Delta_{04}(g_\ast)  \simeq 3.35$. 

Based on the values of $\Delta_{24}(g)$ and $\Delta_{04}(g)$, there can be three situations for a given $g$: 
\begin{enumerate}
\item[i)]  When $c<\text{min}\Big(\Delta_{24}(g), \Delta_{04}(g)\Big)$: the ground state is a $SU(2)_R$ singlet having  $\langle Q_0 \rangle=4$. We refer to this as phase-I.  
\item[ii)]  When $c>\Delta_{04}(g)$:  the ground state is a $SU(2)_R$ singlet with  $\langle Q_0 \rangle=0$. We refer to this as phase-II. 
\item[iii)]  When $\Delta_{24}(g)<c <\Delta_{04}(g)$:  the ground state is a $SU(2)_R$ triplet with  $\langle Q_0 \rangle=2$. This  corresponds to phase-III.  
\end{enumerate}
It is easy to see that phases I and II exist for any $g$, if $c$ is appropriately chosen. However phase III can  be realized only if $(\Delta_{24}-\Delta_{04})$ is negative, which requires $g <g_\ast$.  If $c>\Delta_{04}(g_\ast) \simeq 3.35 $, varying $g$ yields a single level crossing QPT between phase I and II at $g_0(c)$, as shown in Fig.~\ref{Fig_sing_trip_crossing}c. In contrast,   when $ c \in (\Delta_{24, min}, \Delta_{04}(g_\ast))\simeq (1.3 ,3.35)$,  
tuning $g$ leads to two QPTs (see Fig.~\ref{Fig_sing_trip_crossing}b): one between phase II and III at $g_1(c)$, and another between phase III and I at $g_2(c)$. Finally, if $c<\Delta_{24, min} \simeq 1.3 $, there is  no level crossing (Fig.~\ref{Fig_sing_trip_crossing}a) and the ground state is in phase I for any $g$.

 \begin{figure}\begin{center}
\includegraphics[width=18cm]{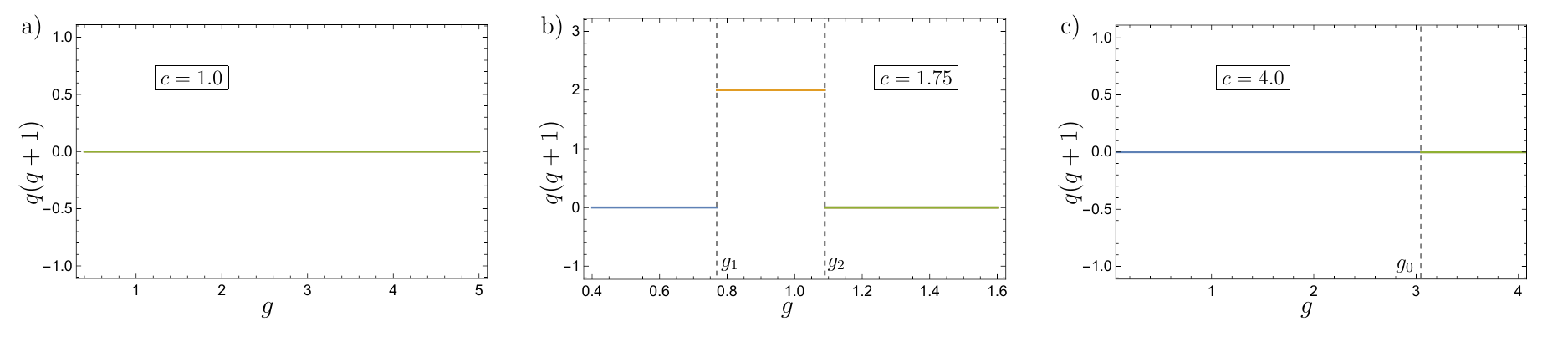}
\caption{  The value of the $SU(2)_R$ Casimir in the ground state as a function of $g$ for  a) when $c=1.0 <\Delta_{24,min}$,  b) when $\Delta_{24,min}<c= 1.75<\Delta_{24}(g_\ast)$, and  c) when $c=4.0>\Delta_{24}(g_\ast)$.   } \label{Fig_casimir_vs_g_diff_c}
 \end{center}
\end{figure}

\begin{figure}\begin{center}
\includegraphics[width=18cm]{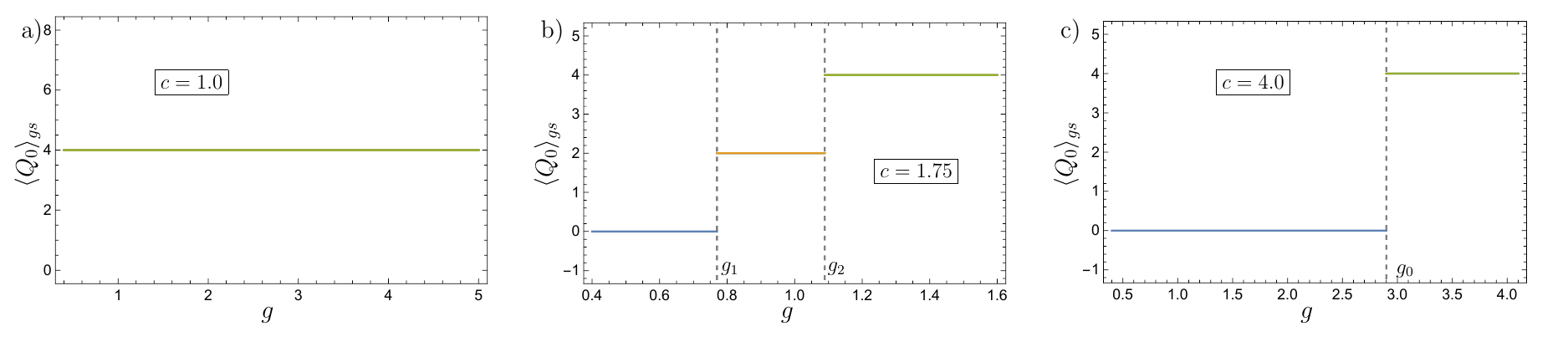}
\caption{$\langle Q_0\rangle_{gs}$ as a function of $g$ for a) when $c=1.0 <\Delta_{24,min}$,  b) when $\Delta_{24,min}<c= 1.75<\Delta_{24}(g_\ast)$, and  c) when $c=4.0>\Delta_{24}(g_\ast)$.} \label{Fig_n_F_vs_g_diff_c}
 \end{center}
\end{figure}

\begin{figure}\begin{center}
\includegraphics[width=18cm]{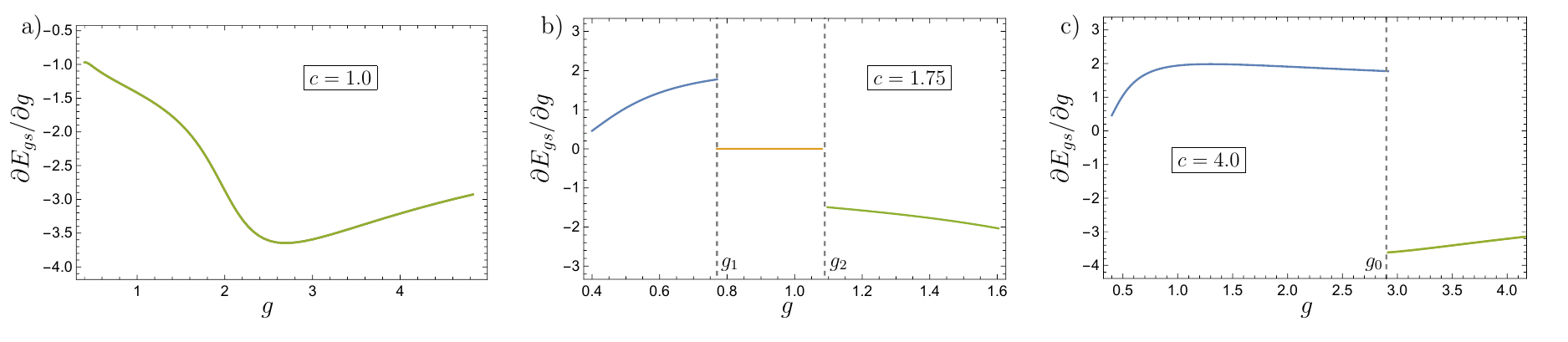}
\caption{ $\partial E_{gs}/\partial g$ as a function of $g$ for a) when $c=1.0 <\Delta_{24,min}$,  b) when $\Delta_{24,min}<c= 1.75<\Delta_{24}(g_\ast)$, and  c) when $c=4.0>\Delta_{24}(g_\ast)$.   } \label{Fig_dEdg_vs_g_diff_c}
 \end{center}
\end{figure}

The curves $g_1(c)$, $g_2(c)$ and $g_0(c)$ denote the phase boundaries where two phases can coexist, and $(g_\ast, \Delta_{04}(g_\ast))\simeq (2.4, 3.35)$ corresponds to the triple point.  At these phase boundaries the quantum numbers of the ground state change, which are captured by various observables. As the ground state on either side of $g_0(c)$ is a $SU(2)_R$ singlet,  the corresponding Casimir  $q(q+1)$ remains continuous across this QPT (Fig.~\ref{Fig_casimir_vs_g_diff_c}c).  In contrast, $g_1(c)$ and $g_2(c)$ are boundaries between a singlet phase and the triplet phase. The value  $q(q+1)$ jumps discontinuously across the QPTs at $g_1(c)$ and $g_2(c)$ (Fig.~\ref{Fig_casimir_vs_g_diff_c}b). Further, the ground state expectation value  $\langle Q_0 \rangle_{gs}$ ( $=\partial E_{gs}/\partial c$) and $\partial E_{gs}/\partial g$ change discontinuously across any of these QPTs (Fig.~\ref{Fig_n_F_vs_g_diff_c} and Fig.~\ref{Fig_dEdg_vs_g_diff_c}).

\begin{figure}[h!]
 \begin{center}
\includegraphics[width=10cm]{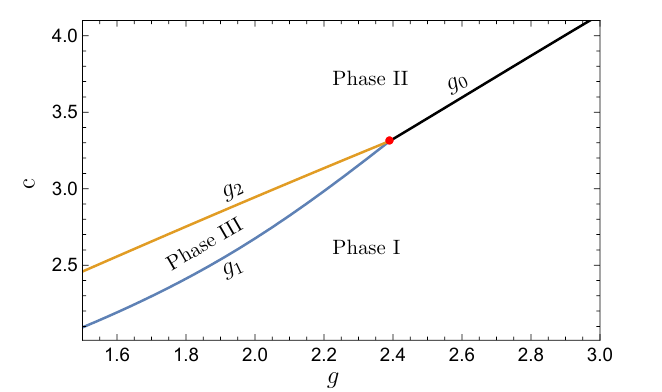}
\caption{The schematic of the phase diagram  in the $g-c$ plane } \label{Fig_phase_diag}
 \end{center}
\end{figure}

The phase diagram in the $g-c$ plane is presented in Fig.~\ref{Fig_phase_diag} and the properties of the phases are summarized in Table \ref{Table_1}.
\begin{table}[H]
\begin{center} {\small 
\begin{tabular}{|c|c|c|c|c|c|l|l|}
\hline &&&&&&\\
Phase & Ground  & Ground   &  $\langle Q_0 \rangle_{gs}$ & $q(q+1)$  & degen-  & \quad\quad\quad Regime of emergence   \\ 
 &  state &  state energy &  &  &eracy  &   \\
  &  
  & $E_{gs}(g,c)$  &  &  &  &  \\
\hline \hline &&&&&& \\
I & $|\Psi_{0s}^{(4)}\rangle$ & $E_{0s}^{(4)}(g,0) + 4 c$ & 4 & 0 & 1 &1. \,\, $c > \Delta_{04}(g_\ast)$ and $g_0(c)<g$ 
\\ &&&&&&\\
 &  &  & && &2. \,\, $\Delta_{24, min}<c< \Delta_{04}(g_\ast)$ 
  \\ 
 &&&&&&\quad\quad and   $g_2(c)<g $ \\&&&&&&\\
  &  &  & && &3. \,\, $\Delta_{24, min}>c$ and   any $g$ \\ &&&&&&\\ \hline  &&&&&&\\
 II & $|\Psi_{0s}^{(0)}\rangle$ & $E_{0s}^{(0)}(g,0) $ & 0 & 0 & 1 &1. \,\, $c > \Delta_{04}(g_\ast)$  and $g_0(c)>g$ \\ 
&&&&&&\\
  & &  & &  &  & 2. \,\, $\Delta_{24, min}<c< \Delta_{04}(g_\ast)$  \\ 
  &&&&&&\quad\quad and   $g_1(c)<g $\\&&&&&&\\ \hline &&&&&&\\
 III & $|\Psi_{0t}^{(2)}\rangle$ & $E_{0t}^{(2)}(g,0) + 2 c$ & 2 & 2 & 3 &  \quad $\Delta_{24, min}<c< \Delta_{04}(g_\ast)$  \\ &&&&&&\quad and   $g_2(c)<g<g_1(c) $\\ &&&&&&\\ \hline 
\end{tabular}
}\\
\mbox{}\\
\caption{The phases for $c>0$ and their properties.}\label{Table_1}
\end{center}
\end{table}

\subsubsection{Spontaneous breaking of $SU(2)_R$ in phase III} 
Interestingly, in the phase III the ground state is a $SU(2)_R$ triplet. This state  is three-fold degenerate and has  Casimir $1(1+1)=2$.  These degenerate states are labelled by the baryon number:  a baryon $|\Psi_{0t}^{(2),+}\rangle$ with $Q_3=1$, a meson $|\Psi_{0t}^{(2),0}\rangle$ with  $Q_3=0$, and an anti-baryon $|\Psi_{0t}^{(2),-}\rangle$ with  $Q_3=-1$.  To see that the $SU(2)_R$ symmetry is spontaneously broken, let us deform the Hamiltonian as 
\begin{equation}
H_{\mu} = H + \mu R^{-1} Q_3, \quad \mu \in \mathbb{R},  
\end{equation}
where $H$ and $Q_p$ are given in (\ref{Ham_1}) and (\ref{SU_2B_charges}), and  $\mu$ is  baryon number chemical potential.  Obviously, $H_{\mu}$ explicitly breaks the $SU(2)_R$ symmetry to a $U(1)_B$  generated by $Q_3$.  

As an algebraic relation, all $Q_p$'s commute with the Hamiltonian only in the $\mu\to 0$ limit, where one might expect the symmetry to be restored. However, this formal restoration does not guarantee that the ground state in the $\mu\to 0$ limit  is  $SU(2)_R$-invariant.  If the limiting ground state depends on the manner in which the limit is taken, the symmetry is spontaneously broken, which is what we will show below.

In the basis $\{|\Psi_{0t}^{(2),+ }\rangle, |\Psi_{0t}^{(2),0}\rangle, |\Psi_{0t}^{(2),- }\rangle \}$, the Hamiltonian  $H_{\mu}$ takes the form
\begin{eqnarray}
H_{\mu} = 
R^{-1} E_{0t}^{(2)}(g,c) I_3+ \mu R^{-1} \left(
\begin{array}{ccc}
1 &0 & 0 \\
0& 0 &0\\
0 & 0& -1
\end{array}\right)
\end{eqnarray}
where $I_3$ is the $3\times 3$ identity matrix.  To ensure that the ground state corresponds to phase III, $g$ and $c$ are suitably chosen in the ranges specified in Table~\ref{Table_1}. The ground state  of $H_{\mu} $ is 
\begin{eqnarray}
|\Psi_{gs}\rangle =\left\{ \begin{array}{ll}
|\Psi_{0t}^{(2),- }\rangle \quad & \text{if } \mu >0 \\ \\
|\Psi_{0t}^{(2),+ }\rangle\quad & \text{if } \mu <0
\end{array}\right. 
\end{eqnarray}
with energy $E_{gs}(g,c, \mu)= E_{0t}^{(2)}(g,c)- |\mu|$.  It is straightforward to see that  $\langle\Psi_{gs}| Q_3 |\Psi_{gs}\rangle= -\text{sgn}(\mu) $.  Importantly,  $|\Psi_{gs}\rangle$ and  $\langle\Psi_{gs}| Q_3 |\Psi_{gs}\rangle$ are both independent of $|\mu|$ and depend only on $\text{sgn}(\mu)$. 

In the $|\mu| \to 0$ limit, the ground state energy smoothly approaches $E_{2t}(g,c)$ irrespective of the sign of $\mu$. However,  $|\Psi_{gs}\rangle$ remains fixed (independent of  $|\mu|$), as long as $\text{sgn}(\mu)$ does not change, and  
\begin{eqnarray}
&& \lim_{\mu \to 0+}|\Psi_{gs}\rangle =|\Psi_{0t}^{( 2) - }\rangle, \quad\quad\quad\quad  \lim_{\mu \to 0-}|\Psi_{gs}\rangle =|\Psi_{0t}^{( 2)+ }\rangle \\ 
&& \lim_{\mu \to 0+}  \langle\Psi_{gs}| Q_3 |\Psi_{gs}\rangle=-1, \quad\quad \lim_{\mu \to 0-}  \langle\Psi_{gs}| Q_3 |\Psi_{gs}\rangle=1. 
\end{eqnarray}
The non-vanishing $\langle\Psi_{gs}| Q_3 |\Psi_{gs}\rangle$ in the  $|\mu| \to 0$ limit implies  that  the $SU(2)_R$ symmetry is spontaneously broken to $U(1)_B$  in phase III.

 \begin{figure}\begin{center}
\includegraphics[width=18cm]{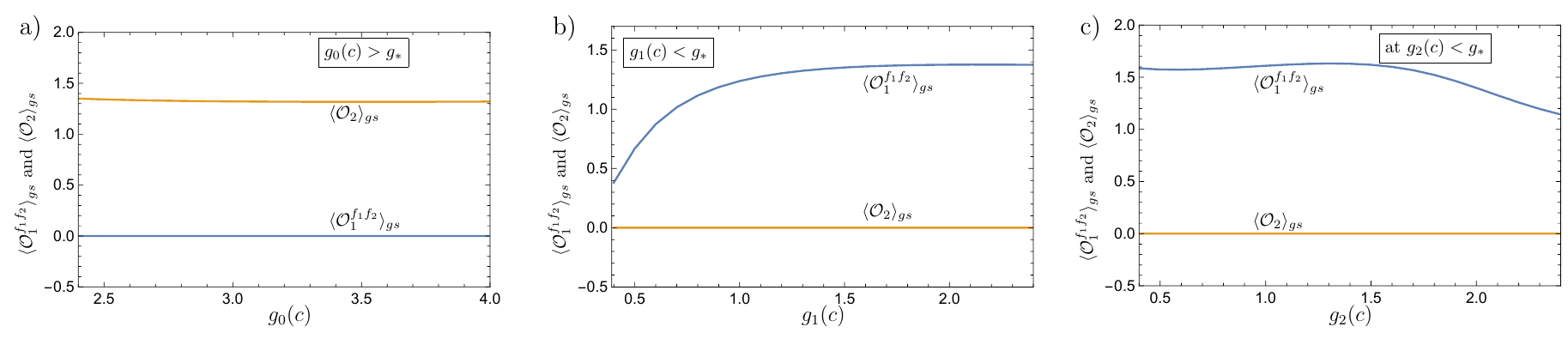}
\caption{  $\langle \mathcal{O}^{f_1,f_2}_1\rangle_{gs}$ and  $\langle \mathcal{O}_2 \rangle_{gs}$ on different phase boundaries a) $g_0(c)$ which separates phases I and II, b) $g_1(c)$ which separates phases II and III, and c)  $g_2(c)$ which separates phases I and III.   } \label{Fig_O1O2}
 \end{center}
\end{figure}
\subsubsection{Phase boundaries} 
Let us turn our attention to the ground state at the phase boundaries.  Let us define the operators (like in \cite{Anber:2018iof}): 
\begin{eqnarray}
&& \mathcal{O}^{(1)}_{f_1,f_2} \equiv  b_{\alpha a f_1}^\dagger \sigma^2_{\alpha \beta} b_{\beta a f_2}^\dagger + b_{\alpha a f_2}\sigma^2_{\alpha \beta}  b_{\beta a f_1}\\
&& \mathcal{O}^{(2)} \equiv  \frac{1}{12}\epsilon_{f_1 f_2} \epsilon_{f_1' f_2'}  b_{\alpha a f_1}^\dagger \sigma^2_{\alpha \beta} b_{\beta a f_2}^\dagger b_{\alpha' a' f_1'}^\dagger \sigma^2_{\alpha' \beta'} b_{\beta' a' f_2'}^\dagger+ h.c 
\end{eqnarray}
where both $ \mathcal{O}^{(1)}$ and $ \mathcal{O}^{(2)}$ transform as  spin-0 color-singlet operators.  Under flavor transformations, $ \mathcal{O}^{(1)}$ is an $SU(2)_R$ triplet while $\mathcal{O}_2$  an $SU(2)_R$ singlet.  Under an axial $\mathbb{Z}_8 \subset U(1)_A$ transformation, both $ \mathcal{O}^{(1)}$  and $ \mathcal{O}^{(2)}$ transform non-trivially:  $ \mathcal{O}^{(1)}$ remains invariant only under a $\mathbb{Z}_2 \subset \mathbb{Z}_8$, while $ \mathcal{O}^{(2)}$ only preserves a $\mathbb{Z}_4 \subset \mathbb{Z}_8$. 

For $c>0$, it is easy to see that  the ground state expectation values of these operators vanish in the bulk of the phases I, II and III.  Thus the bulk ground states preserve the axial $\mathbb{Z}_8$ symmetry.

To study the fate of the axial symmetry on the phase boundaries, we deform the Hamiltonian as 
\begin{eqnarray}
H_\omega = H + \omega R^{-1}\Big(\mathcal{O}^{(1)}_{12}+  \mathcal{O}^{(2)}\Big), \quad\quad \omega \in \mathbb{R},
\end{eqnarray}
which explicitly breaks $SU(2)_R \times \mathbb{Z}_8$  when $|\omega|>0$. The symmetry is formally restored at the phase boundaries and we study the limit $\omega \to 0$ here.

We find that at the phase boundary $g_0(c)$:
\begin{eqnarray}
\langle \mathcal{O}_{12}^{(1)} \rangle_{gs} \equiv \lim_{\omega \to 0} \langle \Psi_{gs}| \mathcal{O}_{12}^{(1)}|\Psi_{gs} \rangle=0, \quad\quad \langle \mathcal{O}^{(2)} \rangle_{gs} \equiv \lim_{\omega \to 0} \langle \Psi_{gs}| \mathcal{O}^{(2)}|\Psi_{gs} \rangle \neq 0,
\end{eqnarray}
as shown in Fig.~\ref{Fig_O1O2}a. This shows that on $g_0(c)$, the axial $ \mathbb{Z}_8$ symmetry is spontaneously broken to $ \mathbb{Z}_4$, while  $SU(2)_R$ remains unbroken.

On the other hand,  at the phase boundaries $g_1(c)$ and $g_2(c)$ (Fig.~\ref{Fig_O1O2}b-c): 
\begin{eqnarray}
\langle \mathcal{O}_{12}^{(1)} \rangle_{gs} \equiv \lim_{\omega \to 0} \langle \Psi_{gs}| \mathcal{O}_{12}^{(1)}|\Psi_{gs} \rangle \neq 0, \quad\quad  \langle \mathcal{O}^{(2)} \rangle_{gs} \equiv \lim_{\omega \to 0} \langle \Psi_{gs}| \mathcal{O}^{(2)}|\Psi_{gs} \rangle= 0. 
\end{eqnarray}
This illustrates that on $g_1(c)$ and $g_2(c)$, the axial $ \mathbb{Z}_8$  and $SU(2)_R$ symmetry are both spontaneously broken to $ U(1)_B \times \mathbb{Z}_2$.

\section{Discussion}\label{sec_4}

In two-flavor matrix adjoint-QC$_2$D, we show that the ground state in intermediate-to-strong coupling is always $SU(2)_R$ symmetric in absence of chiral chemical potential and hence, in such situations the chiral symmetry  remains unbroken. Our analysis reveals two distinct coupling regimes. In the intermediate coupling regime, the ground state is a unique   $SU(2)_R$ singlet. In the strong coupling regime, however, it becomes doubly degenerate, consisting of two  $SU(2)_R$ singlets.  
This degeneracy spontaneously breaks the axial $\mathbb{Z}_8$ (resulting from the anomalous breaking of $U(1)_A$) to a discrete $\mathbb{Z}_4$. Interestingly, this nature of the ground state, its degeneracy and symmetry in the strong coupling regime are in excellent agreement with the effective field theory predictions in \cite{Anber:2018iof}.  

The situation changes dramatically when a chiral chemical potential is added. The resulting level crossings between states with different quark contents are identified as QPTs.
Each phase is characterized by a distinct ground state and its properties.   We have mapped out the full $g-c$ phase diagram, identifying three distinct phases. Two of these phases preserve $SU(2)_R$, while the third one spontaneously breaks $SU(2)_R \to U(1)_B$.  We derived the precise conditions for each phase and found that the $SU(2)_R$-broken phase only exists in a narrow window in the $g-c$ plane, bound by the triple point at $(g,c) \simeq (2.4, 3.35)$. In all these phases, the axial $\mathbb{Z}_8$ remains intact. This axial symmetry is broken only on the phase boundaries. On the coexistence curve between two phase with unbroken $SU(2)_R$, the axial symmetry is spontaneously broken to $\mathbb{Z}_4$. As there are two distinct cosets $\mathbb{Z}_8/\mathbb{Z}_4$, the ground state in this case two-fold degenerate.  In contrast, on the boundary between a $SU(2)_R$-preserving and a $SU(2)_R$-breaking phase, the axial symmetry is spontaneously broken to $\mathbb{Z}_2$. The four-fold degenerate ground state accounts for the index of $\mathbb{Z}_2$ in $\mathbb{Z}_8$.

For larger values of $g$, the qualitative picture is expected to remain more or less the same.  Nevertheless, the strict  $g \to \infty$ limit requires separate consideration, as emergent flat directions may significantly complicate  the ground state structure and the nature of the spectrum. Similarly, the extrapolation to the weak-coupling regime is not straightforward, since localization-delocalization transitions may affect one or more flavor sectors. The fate of chiral symmetry in these regimes remains an important open question and is currently under investigation.

\appendix

\section*{Appendices} 

\section{Symmetries of the Hamiltonian}\label{app_sym}

 The glue Hilbert space $\mathcal{H}_G=\displaystyle{L^2(M_3(\mathbb{R}),\prod_{ia} dM_{ia})}$ is infinite dimensional and the glue states may be organized in representations of spin and color.  The quark Hilbert space $\mathcal{H}_F$ is  $2^{12}$-dimensional, and the quark states can have integer or half-integer spin $0\leq s \leq 3$ which   transform in singlet, triplet or quintuplet representations of color $SU(2)$. The details of these symmetries are discussed below.  The total Hilbert space is $\mathcal{H}_G \otimes \mathcal{H}_F$, where  the states may also be labelled by spin and color.

The glue degrees of freedom transform as spin-1 under spatial rotations and as in the adjoint representation of the $SU(2)$ gauge group. In the   glue Hilbert space $\mathcal{H}_G$, the spatial and gauge rotations are generated by $ \mathcal{L}_i^{\text{glue}}$ and $G_a^{\text{glue}}$, respectively: 
 \begin{eqnarray}
 \mathcal{L}_i^{\text{glue}} \equiv  -\epsilon_{ijk}  P_{ja} M_{ka}, \quad \quad G_a^{\text{glue}}\equiv -\epsilon_{abc} P_{ib} M_{ic}
 \end{eqnarray}
 which satisfy 
 \begin{eqnarray}
 [ \mathcal{L}_i^{\text{glue}},  \mathcal{L}_j^{\text{glue}}]=i\epsilon_{ijk}  \mathcal{L}_k^{\text{glue}}, \quad\quad [G_a^{\text{glue}}, G_b^{\text{glue}}]=i\epsilon_{abc} G_c^{\text{glue}}. 
 \end{eqnarray}
 
 On the other hand, the adjoint quarks transform as spin-$\frac{1}{2}$ under spatial rotations and in the adjoint representation of the gauge group. In the quark Hilbert space $\mathcal{H}_F$, the spatial and gauge rotations are generated by $ \mathcal{L}_i^{\text{quark}}$ and $G_a^{\text{quark}}$, respectively: 
 \begin{eqnarray}
   \mathcal{L}_i^{\text{quark}} \equiv  \frac{1}{2} b^\dagger_{\alpha a f} \sigma^i_{\alpha \beta } b_{\beta a f} , \quad \quad G_a^{\text{quark}}\equiv -i \epsilon_{abc} b^\dagger_{\alpha bf}  b_{\alpha cf} 
 \end{eqnarray}
 which satisfy 
 \begin{eqnarray}
 [ \mathcal{L}_i^{\text{quark}},  \mathcal{L}_j^{\text{quark}}]=i\epsilon_{ijk}  \mathcal{L}_k^{\text{quark}}, \quad\quad [G_a^{\text{quark}}, G_b^{\text{quark}}]=i\epsilon_{abc} G_c^{\text{quark}}. 
 \end{eqnarray}

  In the total Hilbert space  $\mathcal{H}=\mathcal{H}_F \otimes \mathcal{H}_G$,  the spatial rotations are generated by $J_i \equiv  \mathcal{L}_i^{\text{quark}}+ \mathcal{L}_i^{\text{glue}}$ 
 and gauge rotations are generated by $G_a \equiv G_a^{\text{quark}}+ G_a^{\text{glue}}$: 
 \begin{eqnarray}
 [J_i, J_j]=i\epsilon_{ijk} J_k, \quad\quad [G_a, G_b]=i \epsilon_{abc} G_c. 
 \end{eqnarray}
 $J_i$ is the total spin which generate spatial rotation group $SO(3)_{rot}$, while  $G_a$'s are the generators of the Gauss' law constraint. 
 
 With the Hamiltonian (\ref{Ham_1}), it is straightforward to check that 
 \begin{eqnarray}
 [H, J_i]=0=[H, G_a]. 
 \end{eqnarray}
 
 The Gauss law constraint demands that all physical observables  commute with $G^a$. This means that the physical Hilbert space $\mathcal{H}_{phys}$ is the color-singlet subspace of $\mathcal{H}$ and any $| \Psi \rangle \in \mathcal{H}_{phys}$ satisfies  $G^a | \Psi \rangle =0$. 

Additionally, the Hamiltonian (\ref{Ham_1}) commutes with $Q_0 \equiv b^\dagger_{\alpha a f} b_{\alpha a f}$ which generates the classical axial $U(1)_A$ symmetry. In the quantum theory, the $U(1)_A$ symmetry is broken by axial anomaly to $\mathbb{Z}_8$ \cite{Acharyya:2021egi}. Despite this,  $Q_0$ is a well-defined operator in $\mathcal{H}_{phys}$ and  the term $(c Q_0)$ in the Hamiltonian is interpreted as chiral chemical potential (but not in the thermodynamic sense) \cite{Braguta:2015owi, Braguta:2016aov, Acharyya:2024pqj}.



\end{document}